# The crossover from collective motion to periphery diffusion for 2D adatom-islands on Cu(111)

Altaf Karim[1], Abdelkader Kara[2], Oleg Trushin[3], and Talat S. Rahman[2,*]

[1] Advanced Energy Technology, Lawrence Berkeley National Laboratory, Berkeley CA94720
[2] Department of Physics, University of Central Florida, Orlando, FL 32816
[3] Institute of Microelectronics and Informatics, Academy of Sciences of Russia, Yaroslavl 150007, Russia

**Abstract**

The diffusion of two dimensional adatom islands (up to 100 atoms) on Cu(111) has been studied, using the self-learning Kinetic Monte Carlo (*SLKMC*) method [1]. A variety of multiple- and single-atom processes are revealed in the simulations, and the size dependence of the diffusion coefficients and effective diffusion barriers are calculated for each. From the tabulated frequencies of events found in the simulation, we show a crossover from diffusion due to the collective motion of the island to a regime in which the island diffuses through periphery-dominated mass transport. This crossover occurs for island sizes between 13 and 19 atoms. For islands containing 19 to 100 atoms the scaling exponent is 1.5, which is in good agreement with previous work. The diffusion of islands containing 2 to 13 atoms can be explained primarily on the basis of a linear increase of the barrier for the collective motion with the size of the island.

The study of adatom and vacancy island diffusion as a function of size has been an important concern for many theorists [2-9]. Because of the inherent differences in the microscopic processes responsible for diffusion and the scaling of its behavior with size, the discussion has naturally bifurcated into those for the larger islands, usually containing more than 10 atoms, and the smaller ones (N<10). An important study of the diffusion of islands of sizes larger than fifteen (N>15), on metal surfaces, was done by Voter [10]. He showed that the diffusion coefficient D obeys the following scaling law with a constant scaling exponent $\alpha$ (= -1.76)

$$D = cN^{\alpha} \qquad (1),$$

where 'N' is the size of the island and c is a constant. The scaling exponent is expected to reflect the intervening atomistic processes responsible for the diffusion [11,12]. Alternatively, the diffusion coefficient of islands diffusion can be written as:

$$D = D_0 \exp(-E_{eff}/k_B T) N^{\alpha} \qquad (2),$$

Where $D_0$ is the pre-exponential factor and $E_{eff}$ is an effective barrier that can be deduced from an Arrhenius plot.

For larger islands, short-range diffusion of the atoms around the periphery, followed by adjustment of island shape, has been proposed to be the dominant mechanism [3,13]. On the theoretical side, Khare *et al.* [14,15] explained island diffusion in terms of the shape fluctuations of the outer boundary, thereby establishing a relation between the macroscopic motion of islands and the atomistic processes responsible for them. In this work, the authors considered various types of single-atom diffusion mechanisms such as periphery diffusion and terrace diffusion. By allowing island diffusion only through the motion of periphery atoms they obtain the scaling exponent $\alpha = 3/2$.

The above scaling law is not valid for small island diffusion [16]. For small adatom islands earlier experimental studies point to a general decrease in mobility with increasing island size, except for some cases of anomalously large mobility [17,18]. In the case of small Ir islands on Ir(111), concerted gliding motion of the island has been reported [13]. Recent theoretical studies of the energetics and dynamics of 1-10 atom Cu islands on Cu(111) have once again highlighted the role of the concerted motion of the island in controlling its diffusion characteristics [19]. For small islands a consistent knowledge of the variation of their mobility with size and the details of the responsible atomistic processes has been established in our previous work [16].

In this paper we show through systematic size-dependent studies of Cu islands on Cu(111) that there is a crossover in size dependence of the processes responsible of island diffusion. We will show that the effective diffusion barriers of small islands are much smaller than those of the large ones. In other words for the smaller islands the effective diffusion barrier scales linearly with the size. On the other hand the effective barrier for large island diffusion is constant, around 0.79eV for Cu islands on Cu(111), with a scaling exponent $\alpha \approx 3/2$. This value of the effective diffusion barrier is closer to the values of high-energy processes, like corner detachments or kink detachments, participating in periphery-dominated mass transport.

Our purpose in this paper is two-fold. First through a size-dependent study of the island diffusion coefficient, we establish the range of sizes of Cu islands that correspond to specific trends in their diffusion coefficient. Second, through detailed analysis of the frequencies of events of atomistic processes and their energy barriers we reveal evidence of the crossover from collective motion to periphery diffusion for Cu adatom-islands on Cu(111).

Our calculations are based on the recently developed Self Learning Kinetic Monte Carlo (*SLKMC*) [1,20] technique in which we have implemented a pattern recognition scheme that assigns a unique label to the particular environment of the diffusing atom up to several neighbors, enabling efficient storage and retrieval of information on activation energy barriers of possible processes that the system may choose to undergo. Provisions are made for automatic calculation of the energy barrier when a process is first identified and storage of the result in a database. These energy barriers are calculated using a simple method, which maps out the total energy of the system as the diffusing entity moves, in small steps, from the initial to the aimed-at final site. During the ensuing energy minimization procedure, all atoms in the system are allowed to relax in all directions, except for the diffusing atom, whose motion is constrained along the reaction path.  Processes involving multiple atoms can thus be revealed naturally. Extensive comparisons of the resulting energy barriers [1] with those obtained using the more sophisticated nudged elastic-band method show only minor differences. The simpler method yields a reduction of almost two orders of magnitude in the time taken to acquire a comprehensive database.  For the calculation of the total energy of the system, interatomic potentials based on the embedded atom method (EAM) [21] are used. The initial step in the simulation is the acquisition of the database. Once it has become stable --i.e. no new processes appear for some time (for the islands under consideration the data base saturated after about ten million KMC steps)-- the system evolves smoothly through atomistic processes of its choice, and statistics are collected for calculating quantities such as the mean square displacement of the center of mass of the island, correlation functions, and the frequencies of the distinctive atomistic processes.

Our calculations here benefited from several strategic simplifications of our original SLKMC code [1]. Here, exchange processes are not considered, as the pattern recognition scheme for them is more complex than the one implemented here for diffusion via atomic hops. In the present study of diffusion of 2D islands on Cu(111), such processes are not expected to play a major role, because their energy barriers are relatively high. Further, they have not been identified experimentally.

Furthermore, we allow island atoms to occupy only fcc hollow sites, as in the original SLKMC code [1]. Because of the high-energy toll, we prevent, in this work, detachment of adatoms from the island; however, an adatom can still go around the corner so that the periphery diffusion mechanism is operational [10,11]. In other respects we expended our inquiry beyond what was provided for in our original code. In the case of islands of smaller sizes, as we have discussed above, concerted motion might dominate the diffusion process [19]. These processes necessarily involve occupation or transit through the hcp sites, although the exact nature of such processes is not known *a priori*. We have thus performed MD simulations at 500 K to identify diffusion processes, which are not collected in the database of the present version of the *SLKMC* code because of its restriction to fcc sites. The MD simulations are performed with the same interatomic potentials as the *SLKMC* simulations. The new processes revealed in the MD simulations were included in the database and its details can be found in our recent work [16].

Using transition state theory, the rate for an atom to hop to a vacant site is given by $r_i = v_i \exp(-\Delta E_i / k_B T)$. Here $\Delta E_i$ is the activation energy barrier, $k_B$ is the Boltzmann constant, and $v_i$ is the attempt frequency or the so-called prefactor. Since most of the thermodynamics of the system is hidden in the prefactor, it should in principle be sensitive to the details of the atomic environment. The prefactors for the various processes can thus be expected to be different. But although the recipe for calculating prefactors is well defined, carrying out these calculations is non-trivial [6]. Recent calculations of the prefactors for concerted island motion containing 2-7 atoms show some variation with size [19] but the effect is not dramatic. In principle it would be preferable to calculate the prefactors for all the processes present in the database. We leave these calculations for the future, and invoke here the often-used assumption of a standard value of $10^{12}$ sec$^{-1}$ for all prefactors [22]. For further efficiency in the KMC algorithm, we have employed the Bortz-Kalos-Lebowitz (BKL) updating scheme [23], which (as recent works have shown [8,24]), allows one to reach macroscopic time scales of seconds or even hours for simulations at, say, room temperature.

As for the model system, we consider a Cu(111) substrate with an adatom island on top. A KMC simulation step begins by placing an ad-atom island of desired size, in a randomly chosen configuration, on the substrate. The system evolves by performing a process of its choice, from the multitude of possible single- or multiple-adatom jumps at each KMC step. We carried out about $10^7$ such steps at 300 K, 500 K, and 700 K. Typically, at 500 K, $10^7$ KMC steps were equal to $10^{-3}$ sec in physical time. The diffusion coefficient of an adatom island is calculated by

$$D = \lim_{t \to \infty} [R_{cm}(t) - R_{cm}(0)]^2 /2dt,$$

where $R_{CM}(t)$ is the position of the center of mass of the island at time $t$, and $d$ is the dimensionality of the system.

Our results for the calculated size dependence of the diffusion coefficients for Cu islands on Cu(111) at 300K and 500K are shown in Fig. 1. For island sizes larger than 13, we clearly observe a crossover region in the behavior of the diffusion coefficients at all temperatures of interest here. For the size window 19 < N < 100, $D$ follows Eq. (2) and we obtain a scaling exponent α that weakly depends on temperature, i.e., 1.49 < α < 1.64, in the above-mentioned range of temperatures. Consistently, we also found a similar type of crossover in the behavior of effective diffusion barriers as function of size, as shown in Fig. 2. These effective diffusion barriers are calculated from Arrhenius plots of island diffusivity. The effective diffusion barriers are constant in the crossover region of island sizes greater than 13 atoms, as shown in Fig. 2. For the larger islands the results are in agreement with the predictions of Eq. (2). The behavior of $D$ for the smaller island sizes N ≤13 (where Eq. (1) is not valid) is interesting. Contrary to the previous work [8,25] we have not found any size dependent oscillations in $D$ in this crossover region. On the other hand, the effective diffusion barrier increases linearly with size up to 13-atom islands in the cases of Cu. Hence, we may define the effective barrier as $E_{eff} = (\zeta N + C)$ where ζ is the slope of the line representing effective diffusion barriers of smaller islands and $C$ is a constant. We find the value of ζ (increase in the energy barrier per atom) to be 0.055 eV with $C = -0.021$ eV. Using these parameters for the linear fit of the effective diffusion barriers we can fit the equation, $D = D_0 \exp(E_{eff}/k_B T)$, on our data of $D$ for small islands. So the behavior of $D$ for smaller island sizes can be easily understood with the fact of linear dependence of the effective diffusion barrier on size.

We turn now into discussion about the microscopic mechanisms for island diffusion. Here we will try to elucidate the crossover on the basis of microscopic mechanisms and their frequencies. For simplification, we divide all microscopic mechanisms into three categories: collective motion, single-atom processes, and multiple-atom processes. We have seen [16] that the energy barriers for collective motion of smaller islands are comparatively lower than those for other types of single atom processes, hence collective motion is favorable for them. The frequencies of events that occur in the KMC simulation, plotted in Fig. 4, verify this conclusion for islands containing up to 8 atoms. In Fig. 4, the frequency of events was multiplied by the factor $\delta=(\Delta CM_i)/(\Delta CM_{max})$, to magnify those processes that play a role in the change of the center of mass position. For smaller islands, single-atom processes do not make a significant change in the center of mass position, whereas collective motion significantly changes it. This factor $\delta$ is also useful in eliminating such non-diffusive processes as island rotation. In Fig. 4 we can clearly see a crossover region at 9- and 10-atom islands. In this region all types of processes have more closely competing energy barriers. For the larger islands having more than 19 atoms the energy barriers for collective motion of the whole island are very high. In this case island-diffusion takes place mostly through single atom processes like corner-detachments, kink-detachments, or corner-rounding, participating in periphery-dominated mass transport. We observe a crossover in size dependence of frequency of events and diffusion coefficients in the regime where the island size is large enough that its barrier of collective motion approaches in a competition with high energy processes participating in periphery mass transport.

In summary, we have performed a systematic study of the crossover from collective-island motion to periphery atom diffusion for Cu adatom-islands on Cu(111), using a self-learning KMC procedure in which the system is allowed to evolve through mechanisms of its choice[1]. Our simulations for large islands (19 to 100 atoms) exhibit size dependence of diffusion coefficients that is of a power law type. The effective barrier for large island diffusion is constant in our results, which is about 0.79eV. This value is closer to the values of high-energy processes, like corner-detachments or kink-detachments that participate in periphery-dominated mass transport. We also confirm that the scaling rule established for large islands is not valid for small-island diffusion. Effective diffusion barriers for these islands are significantly smaller than those of large islands. In contrast to the case of the large islands, which have almost constant effective diffusion barriers, the small islands show a trend of increasing effective diffusion barrier with island size. In these cases, processes representing the collective motion of the whole island are responsible for small-island diffusion because of their comparatively low energy barriers. We observe a crossover in size dependence of diffusion coefficients in the regime where the island size is big enough that its barrier of collective motion comes into close competition with high-energy processes participating in periphery mass transport.

**Acknowledgements**

We thank Tapio Ala Nissila for helpful discussions and Lyman Baker for careful reading of the manuscript and many constructive comments. This work was supported in part by DOE Grant DE-FG02-07ER46354.

**FIGURE CAPTIONS:**

**FIG. 1**: Adatom island diffusion coefficient D vs N for $1 < N < 101$ at T= 300K and 500K.

**FIG. 2**: Effective diffusion barriers as a function of the island size.

**FIG. 3**: Fit of the proposed theoretical model on simulation data of D.

**FIG. 4**: Frequencies of different processes as function of the island size.

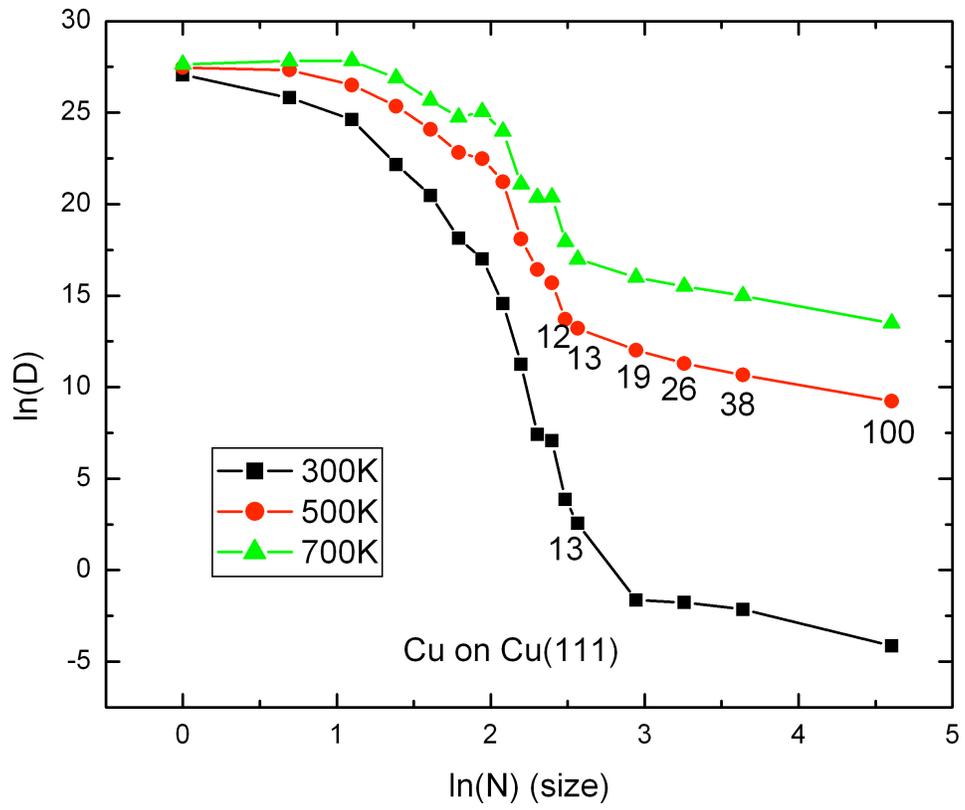

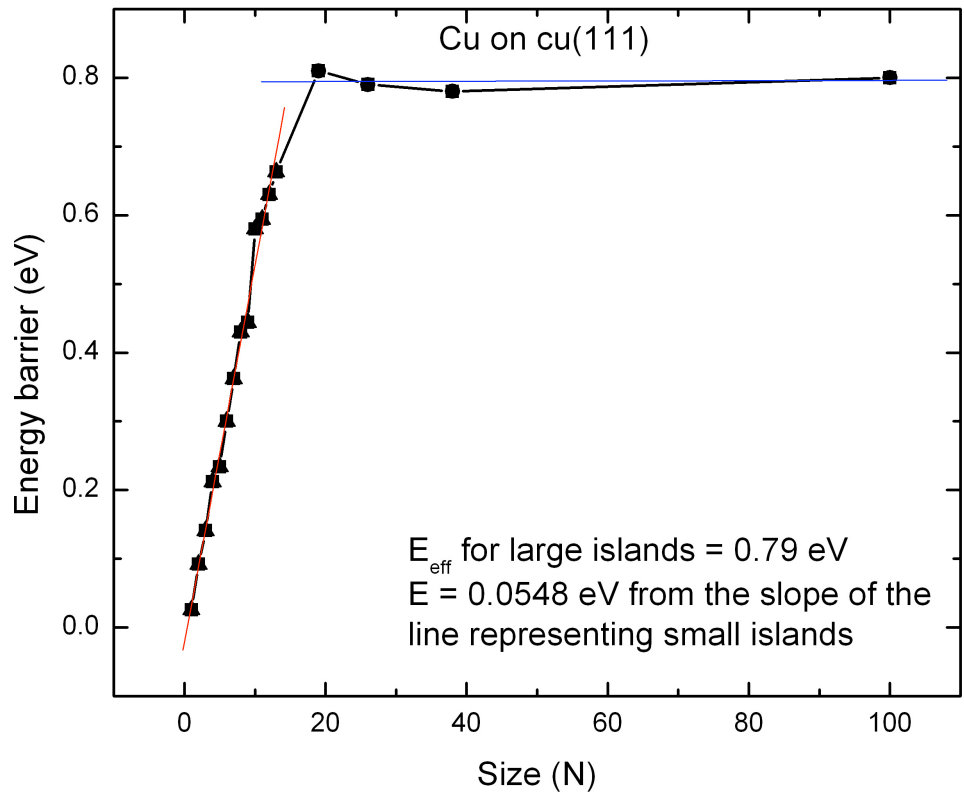

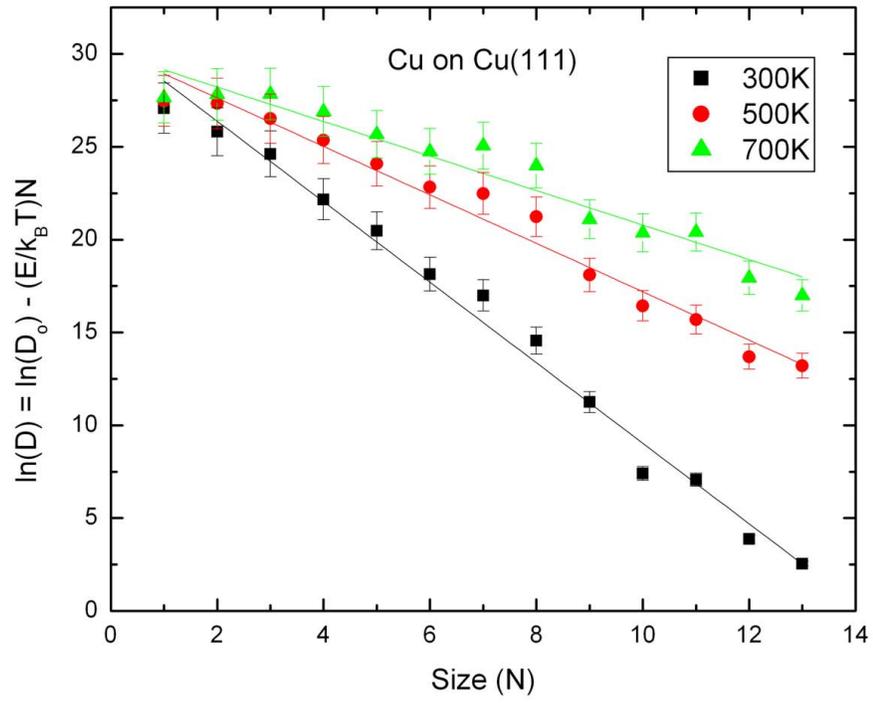

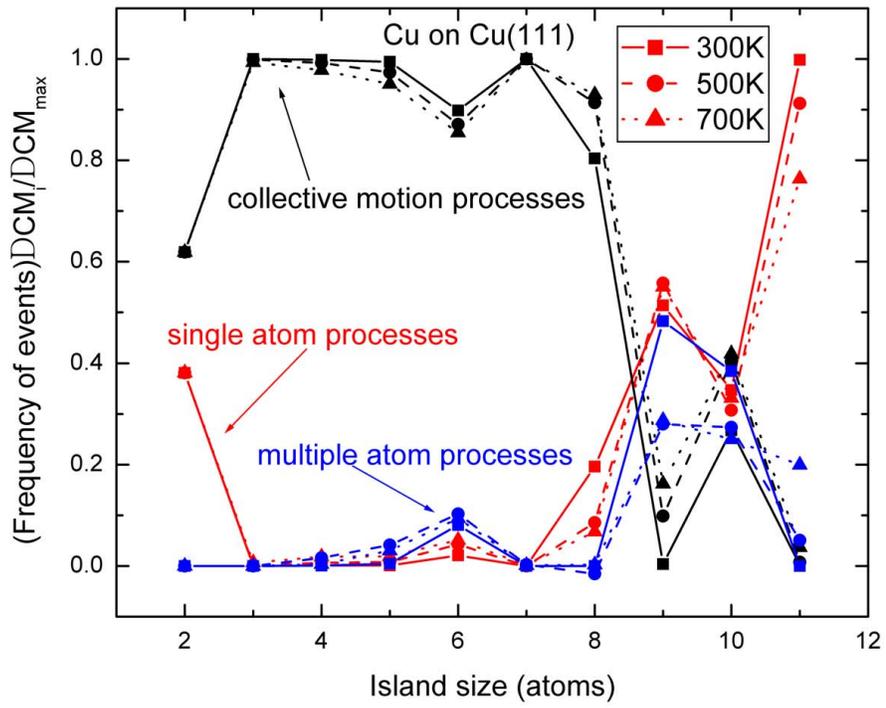